# Itinerant to relocalized transition of $f$ electrons in the Kondo insulator CeRu$_4$Sn$_6$

Fan-Ying Wu,[1] Qi-Yi Wu,[1] Chen Zhang,[1] Yang Luo,[1] Xiangqi Liu,[2] Yuan-Feng Xu,[3]
Dong-Hui Lu,[4] Makoto Hashimoto,[4] Hao Liu,[1] Yin-Zou Zhao,[1] Jiao-Jiao Song,[1] Ya-Hua
Yuan,[1] Hai-Yun Liu,[5] Jun He,[1] Yu-Xia Duan,[1] Yan-Feng Guo,[2, 6, *] and Jian-Qiao Meng[1, †]

[1]*School of Physics and Electronics, Central South University, Changsha 410083, Hunan, China*
[2]*School of Physical Science and Technology, ShanghaiTech University, Shanghai 201210, China*
[3]*Center for Correlated Matter and School of Physics, Zhejiang University, Hangzhou 310058, China*
[4]*Stanford Synchrotron Radiation Lightsource, SLAC National Accelerator Laboratory, Menlo Park, California 94025, USA*
[5]*Beijing Academy of Quantum Information Sciences, Beijing 100085, China*
[6]*ShanghaiTech Laboratory for Topological Physics, ShanghaiTech University, Shanghai 201210, China*
(Dated: Thursday 20$^{\text{th}}$ April, 2023)

The three-dimensional electronic structure and the nature of Ce 4$f$ electrons of the Kondo insulator CeRu$_4$Sn$_6$ are investigated by angle-resolved photoemission spectroscopy, utilizing tunable photon energies. Our results reveal (i) the three-dimensional $k$-space nature of the Fermi surface, (ii) the localized-to-itinerant transition of $f$ electrons occurs at a much high temperature than the hybridization gap opening temperature, and (iii) the "relocalization" of itinerant $f$-electrons below 25 K, which could be the precursor to the establishment of magnetic order.

## I. INTRODUCTION

From a spectroscopic point of view, the most typical feature of heavy fermion (HF) is that as the temperature decreases, the conduction electrons hybridize with localized $f$-electrons ($c$-$f$), forming narrow hybridization gaps and producing various ground states depending on the location of Fermi energy ($E_F$) [1]. If the $E_F$ is located in the gap, the Kondo insulating ground state appears; otherwise, the $E_F$ is situated in one of the hybridized bands, and the metallic HF ground state appears [2, 3]. Metallic HF systems have been studied in depth over the past few decades. In contrast, the physical properties of Kondo insulators are rarely studied, and further research is needed.

Kondo insulators are also known as HF semiconductors. Few HF semiconductors have been studied recently, such as SmB$_6$ [4–12], Ce$_3$Bi$_4$Pt$_3$ [1, 13, 14], and CeNiSn [16, 17]. Notably, the band topologies of Kondo insulators have attracted extensive interest because, theoretically, strongly correlated interactions can produce topologically insulating phases such as SmB$_6$ [4–6] and Ce$_3$Bi$_4$Pt$_3$ [13, 14]. More careful study reveals that some non-cubic structured Kondo insulator systems, such as CeRhSb and CeNiSn [15–17], have strong anisotropic hybridization amplitude. It was considered to be rather a Kondo semimetal [18]. Hybridization gaps with nodes [19] or V-shaped DOS [20] or external influences, including stoichiometry and impurity [21], have served as possible explanations. The scarcity of research subjects limits the study of HF semiconductors or Kondo semimetals.

The low-carrier density HF compounds CeRu$_4$Sn$_6$ crystallizes in the body-centered tetragonal structure of space group $I\bar{4}2m$ with no inversion center ($a$ = 6.8810 Å, $c$ = 9.7520 Å) [22–27]. Since $c/a \approx \sqrt{2}$, $c' \approx c$, where $c'$ is the diagonal of the tetragonal plane, differs by only 0.2% from $c$. Thus, CeRu$_4$Sn$_6$ can also be treated as a quasi-cubic structure [24]. Three heavy atomic elements guarantee strong spin-orbit interaction, resulting in the splitting of the $4f^1$ state into $4f^1_{5/2}$ and $4f^1_{7/2}$ [28]. X-ray absorption spectroscopy determined a high Kondo temperature $T_K \approx$ 170 K [29]. Resistivity measurements show that it exhibits the semimetallic character [30]. Evidence of the onset of an energy gap at about 30 K has been found in the temperature dependence of several quantities: electrical resistivity and thermal conductivity [23, 31], thermopower [31], and spin-lattice-relaxation rate in nuclear-magnetic-resonance (NMR) experiments [32]. Other resistivity measurements indicated that the opening temperature of the hybridization gap is of the order of 100 K [22, 24]. Its optical [25], electronic transport [22, 31], and magnetic [24] properties are strongly anisotropic, probably due to its anisotropic gap. At present, no evidence of long-range or even short-range magnetic order has been found at low temperatures [23, 33]. Thermodynamic properties suggest that CeRu$_4$Sn$_6$ is quantum critical without tuning [30]. Theoretical calculation suggested that CeRu$_4$Sn$_6$ may be a Kondo insulator with a gap through most areas of the Brillouin zone [24, 25, 29], a Kondo semimetal with bands crossing at the $E_F$ [34] or Heavy Weyl semimetal with Weyl points in the heavy quasiparticle bands [35]. Currently, the topological properties of the CeRu$_4$Sn$_6$ band are mainly limited to theoretical calculation. To understand its HF physics and topological properties, experimental measurements such as Angle-resolved photoemission spectroscopy (ARPES), which can obtain the band information, are urgently needed.

Here we report ARPES results, using tunable synchrotron radiation, on high-quality single crystals of CeRu$_4$Sn$_6$. The Fermi surface (FS) topology of CeRu$_4$Sn$_6$ was mapped out by systematic photon energy dependence (along the Γ-X direction) and constant photon energy ($h\nu$ = 80 and 85 eV) ARPES measurements. The measured FS topology is compared with LDA + Gutzwiller calculations [35]. The itinerant-to-localized transition of Ce 4$f$ electrons was investigated by employing temperature-dependent Ce 4$d$-4$f$ on-resonance ARPES measurements.

## II. EXPERIMENTAL

ARPES measurements were carried out at beamline 5-2 of the Stanford Synchrotron Radiation Lightsource (SSRL) using a Scienta DA30L electron spectrometer. All samples were cleaved *in situ* and measured in an ultrahigh vacuum with a base pressure better than $6 \times 10^{-11}$ mbar. CeRu$_4$Sn$_6$ FS topology along the Γ-X (perpendicular) direction in the high-symmetry Γ-X-P-Z plane was mapped out in the photon energy range 40 to 90 eV in steps of 1 eV, with varying energy resolution (8-18 meV). Photon energies of 80 and 85 eV with an energy resolution of ∼ 16 meV were chosen to probe FSs at the $k_x$-$k_z$ plane. Off- and on-resonance spectra ($h\nu$ = 114 and 121/122 eV, respectively) were measured (110) surface along $\bar{Z}-\bar{\Gamma}-\bar{Z}$ direction with a total energy resolution ∼ 20 meV to investigate the nature of Ce 4$f$ electrons. Detailed temperature-dependent of on-resonance ARPES measurements were conducted to reveal the itinerant-to-localized transition of Ce 4$f$ electrons.

The CeRu$_4$Sn$_6$ crystals were synthesized by using Pb as the flux. High-purity elements of Ce powder (Aladdin, 99.5%), Ru power (Adamas, 99.9%), Sn pellet (Macklin, 99.99%), and Pb granular (Macklin, 99.99%) were mixed in a molar ratio of 1:4:6:80 and placed into an alumina crucible which was then sealed into a quartz tube in a vacuum. The assembly was heated in a furnace up to 1150 °C within 20 hrs, kept at this temperature for 20 hrs, and then slowly cooled down to 650 °C at a temperature-decreasing rate of 2 °C/h. The excess Pb was removed at this temperature by quickly placing the assembly into a high-speed centrifuge.

## III. RESULTS AND DISCUSSION

The bulk Brillouin zone (BZ) of CeRu$_4$Sn$_6$, as well as their projected surface BZs for both (100) and (001) surfaces, are shown in Fig. 1(a). Figure 1(b) shows the calculated three-dimensional (3D) FS of CeRu$_4$Sn$_6$ [35]. FSs show rather strong 3D characters. All FSs are close to the bulk BZ center Γ point, including the tiny FSs wrapped in purple FS that are not shown. Figures 1(c1) and 1(d1) display the calculated two-dimensional FS contours at the Γ-X-P-Z plane and $k_x$-$k_z$ plane ($k_y$ = 0), respectively. As shown by the solid lines and dashed lines, the calculated constant energy contours of CeRu$_4$Sn$_6$ varies remarkable with the small change of the binding energy $E_B$. The FS sheets in green are the tiny pockets not shown in Fig. 1(b).

Figure 1(c2) shows experimental FS topologies of CeRu$_4$Sn$_6$ obtained from photon energy-dependent normal emission of (110) surface at a temperature of 20 K. The measurement was performed in a section of the high-symmetry plane spanned by (1,1,0) and $k_z$. Different $k_\parallel$ values were accessed by varying photon energies between 40 and 90 eV. The corresponding momentum range covers more than a full BZ and includes both Γ and X points. Comparing the theoretical calculations with the measured intensities shows a qualitative agreement between them. Fermi sheets' intensity and shapes vary significantly with photon energy. This demonstrates the strong 3D character of the electronic structure of CeRu$_4$Sn$_6$ along the Γ-X direction, consistent with theoretical calculations. It also clarifies that the incident photons probe the bulk band dispersion. Figure 1(d2) displays the constant photon energy $k_x$-$k_z$ [(010) surface] FS maps measured at 14 K with 80 (left panel) and 85 eV (right panel) photon energies. The two energies obtained similar FS topology. A comparison between experimental and theoretical results [Figs. 1(b) and 1(d1)] indicate that these two photon energies measure the plane near the Γ point. The FSs qualitative agree with the calculations. It is worth noting that we have not yet obtained clear FS sheets due to the lack of clear dispersion and the sizeable energy gap through most of the BZ [29, 34], which will be discussed later. We also note that the remarkable 3D electronic structure of CeRu$_4$Sn$_6$ and the significant momentum-broadening effect perpendicular to the surface produced by low photon energies [36–40] will lead to dispersion blurring, which is exactly the case here.

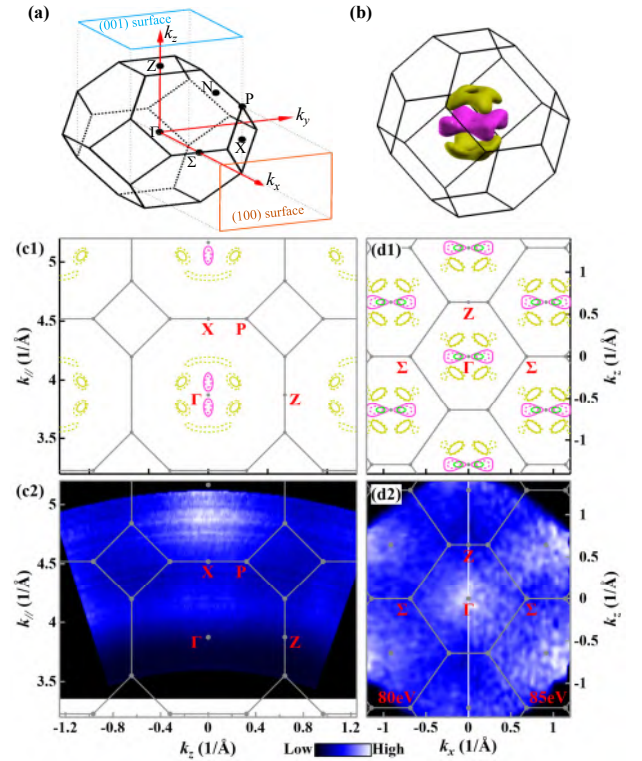

FIG. 1. Fermi surface mappings of CeRu$_4$Sn$_6$ at 20 K. (**a**) The bulk BZ and the projected surface BZ for both (100) and (001) surfaces with high-symmetry momentum points marked. (**b**) Calculated FS of CeRu$_4$Sn$_6$ with LDA + Gutzwiller method. The 3D FS is presented in the body-centered tetragonal Brillouin zone, in which one unit cell contains one Ce atom. Here different color stands for different FS sheets. (**c1**) and (**d1**) Calculated FS contours at the Γ-X-P-Z plane and $k_x$-$k_z$ plane (Γ point), respectively. The solids lines represent the FS contours at $E_F$, and the dashed lines represent the FS contours at $E_B$ = 0.01 eV. (**c2**) Experimental 3D FS map of the $E_F$ intensity constructed out of 51 $E_F$ cuts along $k_z$ through the zone center measured (110) surface with $h\nu$ = 40-90 eV photons in 1 eV steps, in the Γ-X-P-Z plane. The inner potential, $V_0$, was estimated as 14 eV. (**d2**) FS mapping at the $k_x$-$k_z$ plane [(010) surface] taken with 80 (left panel) and 85 eV photons (right panel). All photoemission intensity data were integrated over an [-10 meV, 10 meV] energy window with respect to the $E_F$.

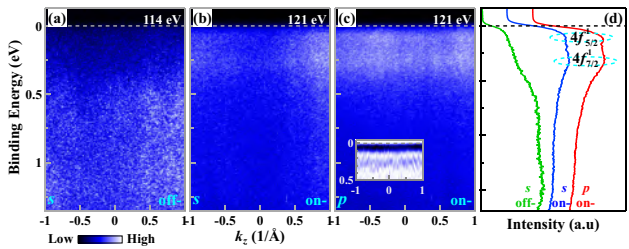

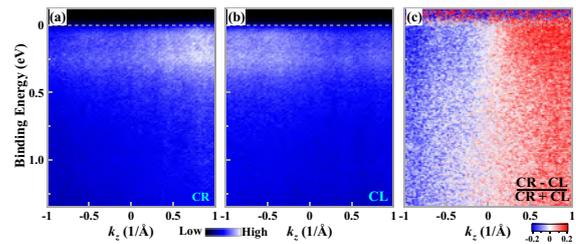

FIG. 2. On- and off-resonance ARPES data of CeRu$_4$Sn$_6$ taken at 14 K with (**a**) off-resonance (114 eV) *s*-polarized, (**b**) on-resonance (121 eV) *s*-polarized, and (**c**) on-resonance *p*-polarized light. Inset: second-derivative image with respect to energy. (**d**) Angle-integrated photoemission spectroscopy of the intensity plot in (a), (b), and (c).

FIG. 3. On-resonance ARPES data of CeRu$_4$Sn$_6$ taken with right- (**a**) and left-polarized (**b**) light. (**c**) The normalized difference [(CR-CL)/(CR+CL)].

Figures 2(a) and 2(b) present off-resonance with 114 eV photons and on-resonance with 121 eV photons photoemission spectra at 14 K, respectively, using *s*-polarized light. Here, the *s* and *p* polarization are perpendicular and parallel to the plane defined by incident light and emitted electrons, respectively. Close to the $E_F$, the off-resonance spectrum shows a density of states of non-*f* orbital character, dominated by Ru 4*d* and Sn 5*p* states derived bands [25, 34]. The Ce 4*f*-electron photoconduction matrix element is strengthened in the on-resonance data, resulting in a clear gathering of the Ce 4*f* spectra weight near the $E_F$. The integral spectra in Fig. 2(d) also show this. It should be noted that the intensity of the conduction band electrons also varies significantly, which is a variation due to the 3D nature of the FS and cannot be ignored here [41]. Two heavy quasiparticle bands originating from the spin-orbit splitting of the $f^1$ final state can be observed in the on-resonance data [Figs. 2(b) and 2(c)], which were commonly observed in other Ce-based HF compounds [38–40, 42–49]. The $f^1_{5/2}$ final state is near $E_F$, and the $f^1_{7/2}$ final state is located at $E_B \sim 0.27$ eV, consistent with the resonant inelastic x-ray scattering measurements [28]. The intensity of the two heavy quasiparticle bands is momentum-dependent, which signifies the *c-f* hybridization. In the on-resonance data, no sharp quasiparticle peaks appear at the $E_F$, which is different from other Ce-based HF compounds such as CePt$_2$In$_7$ [38, 42], Ce$M$In$_5$ ($M$ = Co, Rh, Ir) [39, 40, 45–47], and CePd$_5$Al$_2$ [48]. It may be because CeRu$_4$Sn$_6$ is a Kondo insulator with a sizeable gap through most of BZ [29, 34], resulting in strong inhibition of $f^1_{5/2}$ intensity at $E_F$. It is consistent with the very low density of states of the 4*f* electron near the $E_F$ [29]. Earlier calculations show that the lowest unoccupied Ce 4*f* state is given by $J_z = |5/2, \pm 3/2\rangle$, not $|5/2, \pm 1/2\rangle$, whose energy is pushed up due to strong hybridization with the Ru 4*d* bands [29]. Furthermore, we found that both the $f^1_{5/2}$ and $f^1_{7/2}$ states are sensitive to polarization. Under *p*-polarized photons [Fig. 2(c)], the $f^1$ electron intensity is more strongly enhanced than that taken with *s*-polarized photons [Fig. 2(b)].

As shown in Fig. 3, we measured circular dichroism (CD) in on-resonant ARPES to gain more information about the orbital symmetry of the bands. The difference in sign for positive and negative *k* manifests the dichroic effect. States possessing similar symmetry properties may have the same dichroic response to changes in circular polarization. The spectral intensity is not uniform along the measured direction, which means that the photoemission matrix element has strong moment dependence. Comparing the two data sets [Figs. 3(a) and 3(b)], one can see a clear difference between the two dispersions taken with different circular polarizations. To see CD more clearly, we subtract CL data from CR data and plot the difference data [CD = (CR-CL)/(CR+CL)] in Fig. 3(c). The CD is as large as about 20% near the $E_F$. The intensity profile of the dispersion is antisymmetric about the vertical axis ($k_z = 0$); the + $k_z$ side is positive while the - $k_z$ side is negative. The spectral intensity of both the $4f^1_{5/2}$ and $4f^1_{7/2}$ state shows similar dichroism compared to other bands, which has been found in other HF compounds such as CeIn$_3$ [49]. The findings indicate that 4*f* bands and other conduction bands have the same orbital symmetry, making CeRu$_4$Sn$_6$ a suitable platform for forming a hybridization gap between bands and producing Kondo ground states in Ce compounds.

To understand the critical issues of HF physics, that is, how the *f*-electrons undergo a localized-to-itinerant transition with temperature, temperature-dependent on-resonance ARPES measurement was performed on CeRu$_4$Sn$_6$ with 122 eV photons. Figures 4(a1)-4(a6) display the evolution of the band structures at selected temperatures. Both $4f^1_{5/2}$ and $4f^1_{7/2}$ states weaken and blur as temperature increases until becoming completely indiscernible at the highest experimental temperature of 230 K. This is consistent with substantial progress of the recent ARPES observations on HFs [40, 42–44] and differs from previous theoretical expectations that heavy electrons appear below the coherence temperature $T^*$ at which *c-f* hybridization begins [50].

Figure 4(b) compares EDCs measured at 7, 100, and 230

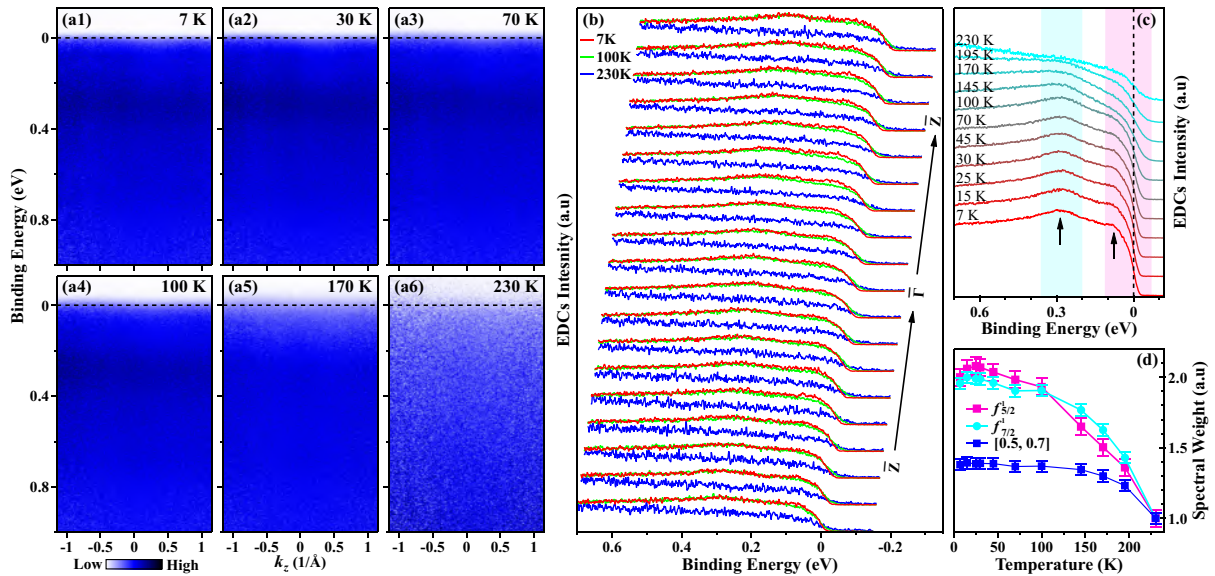

FIG. 4. Temperature evolution of the heavy quasiparticle bands. (**a**) On-resonance (122 eV) band structure of CeRu$_4$Sn$_6$ at labeled temperatures. (**b**) Detailed ARPES spectral of CeRu$_4$Sn$_6$ measured at 7, 100, and 230 K. Arrow indicates angle increase. (**c**) The angle-integrated EDCs over the angle range shown in (a) at various temperatures. (**d**) Temperature dependence of the heavy quasiparticle spectral weight. The cyan and magenta lines represent spectral weight integrals of the light cyan and light pink regions shown in (c), respectively. The blue squares represent the spectral weight integrals over [0.5 eV, 0.7 eV].

K. The 7 K and 100 K spectra are similar, significantly different from those of 230 K. At low temperatures, the heavy $f$-bands span the measured momentum range. In contrast, at high temperatures of 230 K, the heavy quasiparticle peaks are unobserved. Figure 4(c) shows the temperature evolution of the integrated EDCs in the measured momentum range shown in Fig. 4(a). The formation of the heavy quasiparticle peaks begins at a temperature of ∼ 195 K, slightly higher than the $T_K$ [29], and prior to the opening of a hybridization gap of the order of 100 K as suggested by resistivity measurement [23–25]. As the temperature decreases, the heavy bands of $4f^1_{5/2}$ and $4f^1_{7/2}$ states gradually become significant, showing a broad temperature range crossover of the localized-to-itinerant transition of $4f$ electrons.

Figure 4(d) quantitatively presents the Ce $4f$-electron spectral weight evolution with temperature. Spectral weight has been normalized to the highest temperature data. The spectral weight of $4f^1_{5/2}$ and $4f^1_{7/2}$ states exhibit similar temperature evolution behavior. As the cooling begins, the spectral weight of $4f^1_{5/2}$ and $4f^1_{7/2}$ states increases with decreasing temperature, which is consistent with previous results [40, 42–44]. However, as the temperature is further reduced below ∼ 25 K, the $f$-spectral weight deviates from the monotonic increase with decreasing temperature and even shows a slight suppression. This indicates that the collective hybridization process begins to reverse, and the Kondo Lattice quasiparticles start to relocalize. To rule out the possibility that this is due to the temperature-dependent background, Fig. 4(d) also shows the temperature dependence of the spectral weight at high binding energies, integrated over [0.5 eV, 0.7 eV]. It can be seen that the background hardly varies with temperature at low temperatures, in contrast to the evolution of the heavy quasiparticle spectral weight. The temperature evolution of the $f$-electron spectral weight is inconsistent with the generally observed monotonic increase in hybridization with decreasing temperature [40, 43, 44].

The "relocalization" phenomenon has been observed in very few other Ce-based HF compounds, e.g., CePt$_2$In$_7$ [42, 51], CeRhIn$_5$ [52], and CeCoGe$_3$ [53]. It has been interpreted as the presence of low-energy crystal electric field (CEF) excitation [42], or a precursor to magnetic order [51, 52], or the competition between magnetic order and Kondo effect [53]. What is the origin of relocalization in CeRu$_4$Sn$_6$? It is unlikely to be caused by the presence of low-energy CEF excitations since the $f^1_{7/2}$ state away from the $E_F$ is also relocalized at low temperatures, as is the $f^1_{5/2}$ state. The opening of an energy gap may suppress the density of state near the $E_F$. However, for the same reason, the "relocalization" phenomenon cannot be caused by the opening of the hybridization gap, although previous electrical transport [31] and NMR [32] measurements suggest the onset of an energy gap was about 30 K, which was consistent with the observed "relocalization" temperature. It is also unlikely to be caused by the adsorption of many molecules on surfaces below 25 K. For CePt$_2$In$_7$ [42] and CeCoGe$_3$ [53], the suppression of the Kondo peak occurred below ∼ 60 K and ∼ 12 K, respectively, rather than around 25 K. The competition between magnetic order and Kondo effect is also proposed as a possible explanation, since the temperature at which the Kondo peak of CeCoGe$_3$ is suppressed coincides with the antiferromagnetic transition tem-perature [53]. However, this is not the case for CeRu$_4$Sn$_6$, which has no magnetic order in the measured temperature range. Finally, we consider the precursor to magnetic order as as a feasible explanation. At low temperatures, the formation of magnetic order inhibits the Kondo effect, leading to the reversal of collective hybridization and the relocalization of heavy quasiparticles [51, 52]. That is, at a temperature between the hybridization temperature and the magnetic ordering transition temperature, the heavy quasiparticles begin to relocalize. Although no long-range or short-range magnetic order has been found in the concentrated Kondo system CeRu$_4$Sn$_6$ at present [23, 33], some studies suggest that CeRu$_4$Sn$_6$ may be near the antiferromagnetic order mediated by RKKY interaction [30]. The origin of "relocalization" deserves further study, and our results are an appropriate starting point.

## IV. CONCLUSION

To conclude, the electronic structure of HF compound CeRu$_4$Sn$_6$ was investigated by ARPES. Our studies suggest that the FS has a strong 3D topology. We directly observed two Ce $4f$-derived nearly flat bands in the on-resonance data, corresponding to the $f^1_{5/2}$ and $f^1_{7/2}$ states. The $f$ electrons begin to evolve into the formation of HF states at a temperature much higher than the onset temperature of collective hybridization. We also observed the "relocalization" of heavy quasiparticles at low temperatures, which may be the precursor of magnetic order. These findings provide critical insight into the understanding of HF physics.

## ACKNOWLEDGEMENTS

This work was supported by the National Natural Science Foundation of China (Grant No. 12074436, and No. 11574402), the Science and Technology Innovation Program of Hunan Province (No. 2022RC3068), and the open project of Beijing National Laboratory for Condensed Matter Physics (Grant No. ZBJ2106110017). Some preliminary ARPES data were taken at the "Dreamline" beamline of the Shanghai Synchrotron Radiation Facility.